\hsize=13.1truecm
\vsize=19.8truecm
\hoffset=0.6truein
\tolerance=10000

\headline{\hfill TH01.1.4}

\font\big=cmbx10 scaled\magstep2

\input epsf
\epsfverbosetrue

\noindent{\big 
A Jordan-Wigner transformation for the $t-J$ and Hubbard 
models with holes.
}

\vskip 20pt

\hskip 1.2truecm S. E. Barnes and S. Maekawa
\vskip 5pt

\hskip 1.2truecm \vbox{\hsize=4.25truein
\noindent Institute for Materials Research, Tohoku University, 
Sendai 980-8577, Japan }

\vskip 20pt
\hskip 1.2truecm Received

\vskip 20pt

\hskip 1.2truecm \vbox{\hsize=4.25truein 
\noindent{\bf Abstract.} \ 
A Jordan-Wigner (JW) transformation for the $t-J$ and Hubbard models
is described.  Introduced are holon and doublon particles for hole and
double occupied sites.  There is only a single spin sector particle. 
Flux tubes occur in a naturel fashion, within a specific gauge, when
the method is adapted to two dimensions.  In order to accommodate three
dimensions ``flux sheets'' are defined.  The adaptation of the method
to the $t-J$-model in the context of high $T_{c}$ is described.
} \vskip 20pt

While the Jordan-Wigner transformation [1] is well known in the
context spin only models such as the Heisenberg or $X-Y$ model it does
not seem to be widely known that a similar approach can be used for
doped systems such as the $t-J$ or Hubbard models and that this
provides a useful alternative formulation of the two dimensional such
models in the context of high $T_{c}$ superconductors.  The purpose of
the present Letter is to describe such a formalism along with some
elementary conclusions which are evident in the basis of this
formalism.

\def\refI{
[1] 
See, e.g., pages 434 and 446 in:
\hfill \break Negele J W and Orland H 1987 {\it Quantum Many-Particle Systems}
(Reading: Adddison-Wesley)
}

Recall first the basic JW transformation in one dimension.  It is
trivial for $S=1/2$ that $\sigma^{+}_{n}$ obey on-site Fermi
commutations rules and that,
$$
f^{\dagger}_{n} = e^{i\pi \sum_{m=1}^{n-1} \hat Q_{m}} \sigma^{+}_{n}
\eqno(1)
$$
obeys such rules even when the site indices are different.  Here $\hat
Q_{n} = f^{\dagger}_{n}f^{}_{n}$, i.e., is the number of particles at
site $n$.  It is then straightforward to show [1] that, e.g., the one
dimensional Heisenberg Hamiltonian ${\cal H} =  J \sum_{n} \vec S_{n}
\cdot \vec S_{n+1}$ becomes,
$$
 {\cal H} 
= {J\over 2} \sum_{n} (f^{\dagger}_{n} f^{}_{n+1} + H.c.) 
+
J \sum_{n} f^{\dagger}_{n} f^{}_{n} f^{\dagger}_{n+1} f^{}_{n+1}
- J \hat N_{\uparrow} \eqno(2)
$$
where, with $\hat N_{\uparrow} = \sum_{n} f^{\dagger}_{n} f^{}_{n}$, 
the last term reflects an effective chemical potential $\mu = J$.

This can, in fact, be generalised to include charge.  Since
$\sigma^{+}_{n}$ is considered to be a creation operator it is
implicit that the vacuum is the down spin ferromagnetic state. 
Creating a $f^{\dagger}_{n}$-fermion implies an up spin at the site
$n$.  In the usual way a $b^{\dagger}_{n}$-boson corresponds to a true
vacuum state $|0\rangle$ without any particle, i.e., to the presence
of a hole, while $d^{\dagger}_{n}$-boson implies that the site $n$ is
in the state $c^{\dagger}_{\uparrow n}c^{\dagger}_{\downarrow
n}|0\rangle$, which is doubly occupied.  The physical operators are
given by,
$$
\sigma^{+}_{n} = e^{-i\pi \sum_{m=1}^{n-1} \hat Q_{m}} f^{\dagger}_{n},
\eqno(3a)
$$
$$
c^{\dagger}_{\uparrow n} =
\left( f^{\dagger}_{n} b^{}_{n} 
+
d^{\dagger}_{n} e^{-i\pi \sum_{m=1}^{n-1} \hat Q_{m}}
\right) e^{i\pi \hat N_{\uparrow}},
\eqno(3b)
$$
$$
c^{\dagger}_{\downarrow n} =
\left( b^{}_{n} e^{-i\pi \sum_{m=1}^{n-1} \hat Q_{m}}   
-
d^{\dagger}_{n} f^{}_{n}
\right) e^{i\pi \hat N_{\uparrow}},
\eqno(3c)
$$
where now $\hat Q_{n} = f^{\dagger}_{n}f^{}_{n} +
b^{\dagger}_{n}b^{}_{n} + d^{\dagger}_{n}d^{}_{n}$ counts all
particles and where there is a {\it constraint\/} that  $Q_{n}
\le 1$, i.e., the particles have a hard core.  It is straightforward
to check that these obey the appropriate commutation rules when the
site indices are different.

\def\refII{
[2] 
Barnes S E, 1976 J. Phys. F: Met. Phys. {\bf 6} 115, 1375;
J. Phys. F: Met. Phys. {\bf 7} 2637; 1980 Adv. Phys. {\bf 30} 801
}

When the the site indices are the same it is {\it not\/} the case that
$\{c^{}_{\uparrow n}, c^{\dagger}_{\uparrow n}\} =1$ if the algebra of
the auxiliary particles is applied without strictly applying the
constraint.  What {\it is\/} easily verified is that the matrix
elements of say $c^{\dagger}_{\uparrow n}$ {\it on the physical
sub-space\/} are correctly given, however it is implied that in the
product, e.g., $c^{\dagger}_{\uparrow n} c^{}_{\uparrow n}$ it is
necessary to include a physical complete set of states between the two
operators.  In this regard the present scheme differs from the
traditional auxiliary particle scheme [2] for which this is not necessary. 
It is implied that care must be exercised when constructing, in
particular, the auxiliary particle version of the Hamiltonian.  Every
physical operator, for the site $n$, can be written in terms of the
product of one or two of the new auxiliary particles since any
operator can be written in terms of:
$$
|\downarrow\rangle \langle \downarrow|
=
1 - (f^{\dagger}_{n}f^{}_{n} +
b^{\dagger}_{n}b^{}_{n} + d^{\dagger}_{n}d^{}_{n}),
\ \ \
|\uparrow\rangle \langle \uparrow|
=
f^{\dagger}_{n}f^{}_{n} 
 \ \ \
 |0\rangle \langle 0|
=
b^{\dagger}_{n}b^{}_{n},
\ \ \
|\uparrow\downarrow\rangle \langle \uparrow\downarrow|
= d^{\dagger}_{n}d^{}_{n}
$$
$$
|0\rangle \langle  \uparrow|
=
b^{\dagger}_{n}f^{}_{n},
\ \ \
| \uparrow  \downarrow \rangle \langle  \uparrow|
=
d^{\dagger}_{n}f^{}_{n},
\ \ \
| \uparrow  \downarrow \rangle \langle  0|
=
d^{\dagger}_{n}b^{}_{n},
\ \ \
|\uparrow\rangle \langle  \downarrow|
=
e^{-i\pi \sum_{m=1}^{n-1} \hat Q_{m}} f^{\dagger}_{n},
$$
$$
|0 \rangle \langle  \downarrow|
=
e^{-i\pi \sum_{m=1}^{n-1} \hat Q_{m}} e^{i\pi \hat N_{\uparrow} } 
b^{\dagger}_{n},
\ \ \
| \uparrow  \downarrow \rangle \langle  \downarrow|
=
e^{-i\pi \sum_{m=1}^{n-1} \hat Q_{m}} e^{i\pi \hat N_{\uparrow}} d^{\dagger}_{n},
\eqno(4)
$$
and their Hermitian conjugates.

For the one dimensional $t-J$-model the phase operators, $e^{i\pi
\hat N_{\uparrow}}$ and $e^{-i\pi \sum_{m=1}^{n-1} \hat Q_{m}} $
cancel and the result is,
$$
{\cal H} = 
- t \sum_{n} (f^{\dagger}_{n} b^{}_{n} b^{\dagger}_{n+1} f^{}_{n+1} 
+ b^{\dagger}_{n} b^{}_{n+1} + H.c.)
+ {J\over 2} \sum_{n}
(f^{\dagger}_{n} f^{}_{n+1} + H.c.)
$$
$$
+ {J\over 4} \sum_{n}
(2f^{\dagger}_{n} f^{}_{n} + b^{\dagger}_{n} b^{}_{n})
(2 f^{\dagger}_{n+1} f^{}_{n+1}+ b^{\dagger}_{n+1} b^{}_{n+1})
- J \hat N_{\uparrow}.
\eqno(5)
$$
The Hubbard model becomes:
$$
{\cal H} = - t \sum_{n} (f^{\dagger}_{n} b^{}_{n} b^{\dagger}_{n+1} f^{}_{n+1} 
+d^{\dagger}_{n} f^{}_{n} f^{\dagger}_{n+1} d^{}_{n+1}
+ b^{\dagger}_{n} b^{}_{n+1}
+ d^{\dagger}_{n} d^{}_{n+1} + H.c.) + U d^{\dagger}_{n} d^{}_{n}
$$
$$
- t \sum_{n} 
(f^{\dagger}_{n} b^{}_{n} d^{}_{n+1} e^{-i\pi \sum_{m=1}^{n} \hat Q_{m}}
+ d^{}_{n} f^{\dagger}_{n+1} b^{}_{n+1} e^{-i\pi \sum_{m=1}^{n-1} \hat Q_{m}}
$$
$$
+ d^{\dagger}_{n} f^{}_{n} b^{\dagger}_{n+1} e^{-i\pi \sum_{m=1}^{n} \hat Q_{m}}
+ b^{\dagger}_{n} d^{\dagger}_{n+1} f^{}_{n+1}  e^{-i\pi \sum_{m=1}^{n-1} \hat Q_{m}}
 + H.c.).
 \eqno(6)
$$

Even in one dimension, for the Hubbard model, the JW formulation would
seem not very useful because of the presence of the factors of
$e^{-i\pi \sum_{m=1}^{n-1} \hat Q_{m}} $.  The situation is even worse
in two dimensions.  The transformation can be used for a two
dimensional system once a mapping to one dimension is defined.  A
possible such map, shown in figure~1a, is to use a zig-zag path whence
even for the $t-J$-model the factors of $e^{-i\pi \sum_{m=1}^{n-1}
\hat Q_{m}} $ appear.  Each site is ordered by a single site
label and the sum in this factor simply counts the number of
particles with smaller such labels.  It is well known that, e.g., the
up spins of the problem, which are equivalent of hard core bosons,
might be converted to fermions by attaching a unite flux tube.  The
awkward factors of $e^{-i\pi \sum_{m=1}^{n-1} \hat Q_{m}} $ in fact
represent such flux tubes a certain ``string'' gauge.

\topinsert \centerline{ \epsfxsize=5.2truein \epsfbox{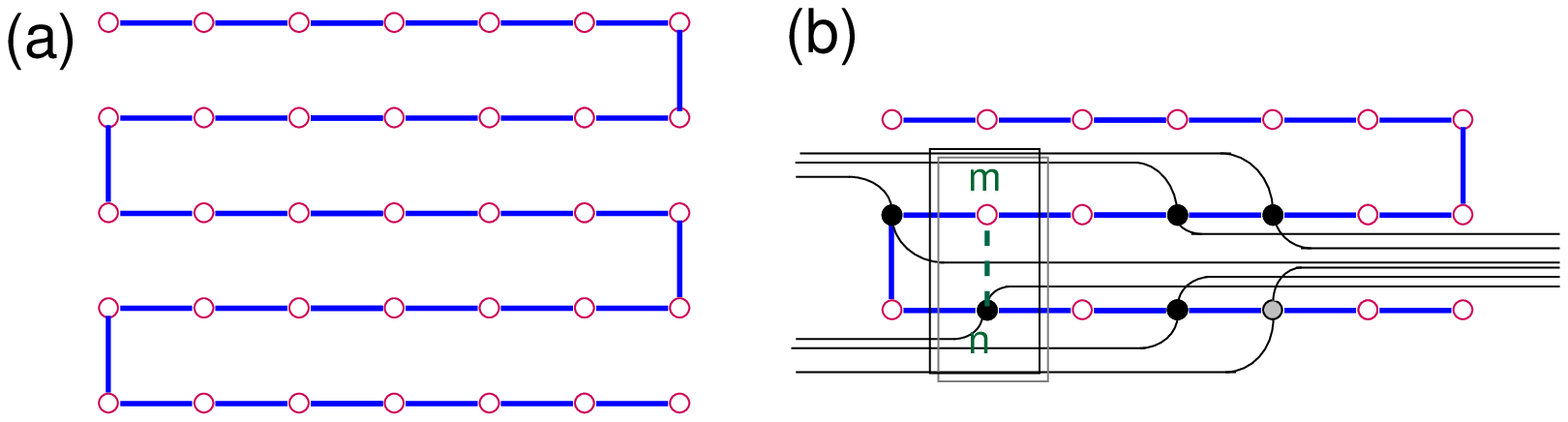} }
\vskip 8pt
\noindent{Fig.  1 (a) The zig-zag path used for the Jordan-Wigner
transformation.  (b) The ``string'' flux lines used to evaluate the
sign changes due to the Jordan-Wigner transformation.} 
\endinsert

Consider the appropriate such phase factor for vertical exchange
$J$-coupling, between sites $n$ and $m$, as shown in figure~1b. The
appropriate term in the Hamiltonian is easily seen to be,
$$
e^{-i\pi \sum_{q=n+1}^{m-1} \hat Q_{q}} J 
(f^{\dagger}_{m} f^{}_{n} + H.c.);
\ \ \ \ n < m,
\eqno(7)
$$
where the sum $\sum_{q=n+1}^{m-1} \hat Q_{m}$ counts the number of
particle between the two sites following the prescribed path.  The
resulting phase factor $e^{-i\pi \sum_{q=n+1}^{m-1} \hat Q_{q}}$ can
be evaluated using the diagramme method illustrated in this figure. 
To each particle are attached two lines or ``strings'', as shown, and
which leave the system without crossing the zig-zag path, i.e., leave
by the open ends.  It follows that,
$$
e^{-i\pi \sum_{q=n+1}^{m-1} \hat Q_{q}} = (-1)^{L},
\eqno(8)
$$
where $L$ is the number of lines which cross a straight line joining
the sites $n$ and $n^{\prime}$.  Put differently, $L$ is the number of
lines cut when the particle hops from site $n$ to site $m$.  In a
fermion plus flux tube formulation the same phase factor would be
$\exp[- i \int_{n}^{m} \vec a \cdot d \vec r]$, where $\vec a$ is an
appropriately defined vector potential.  The identification is
therefore:
$$
\int\vec a \cdot d \vec r = \pi L.
\eqno(9)
$$
Any path which encloses a single particle cuts two lines and has,
$$
\int\vec a \cdot d \vec r = 2\pi,
\eqno(10)
$$
corresponding to a single flux quantum.  Evidently different mappings
from one to two dimensions corresponds to a different version of this
JW string gauge.  The phase factors appropriate to correlation
function are also determined, e.g.,
$$
\langle [S^{+}_{n}(t),S^{-}_{m}(0)]\rangle
=
(-1)^{L_{nm}}\langle 
[f^{\dagger}_{n}(t),f^{}_{m}(0)]\rangle,
\eqno(11)
$$
where $L_{nm}$ is the number of lines crossed for any path joining 
the sites $n$ and $m$ which does not pass through any particle. (It 
need not necessarily follow bonds.) 

With this method, e.g., the full $t-J$-model for a two dimensional 
square lattice is,
$$
{\cal H} = 
-  \sum_{<ij>}  (t f^{\dagger}_{i} b^{}_{i} b^{\dagger}_{j} f^{}_{j} 
+ t_{ij} b^{\dagger}_{i} b^{}_{j} + H.c.)
+  \sum_{<ij>}{J_{ij}\over 2}
(f^{\dagger}_{i} f^{}_{j} + H.c.)
$$
$$
+ {J\over 4}\sum_{<ij>}
(2f^{\dagger}_{i} f^{}_{i} + b^{\dagger}_{i} b^{}_{i})
(2 f^{\dagger}_{j} f^{}_{j}+ b^{\dagger}_{j} b^{}_{j})
- J (2 \hat N_{\uparrow}+ \hat N_{0}), 
\eqno(12)
$$
where $t_{ij} = (-1)^{L_{ij}}t$ and $J_{ij} = (-1)^{L_{ij}}J$ and 
where $L_{ij}$ is the number of strings crossed in going for the site 
$i$ to $j$ as described above. Here $\hat N_{0} = \sum_{i} 
b^{\dagger}_{i} b^{}_{i}$. There is no factor $(-1)^{L_{ij}}$ in 
the $f^{\dagger}_{i} b^{}_{i} b^{\dagger}_{j} f^{}_{j}$ term since the
factors $e^{-i\pi \sum_{m=n+1}^{n^{\prime}-1} \hat Q_{m}}$ cancel.  
(The term involves the exchange of particles and hence each crosses 
the same number of strings and the total number is necessarily even.)  
These factors also cancel in static interaction term with the 
prefactor ${J\over 4}$.  (Here there is no movement of particles and 
hence no strings are crossed.)

It is rather obvious, for the, $x=0$, undoped case, that for the physical
correlation function $\langle [S^{+}_{n}(t),S^{-}_{m}(0)]\rangle =
(-1)^{L_{nm}}\langle[f^{\dagger}_{n}(t),f^{}_{m}(0)]\rangle$ the
string prefactor $(-1)^{L_{nm}}$ cancels against the similar factors
in $J_{ij}$ contained in the propagator and that hence this quantity
is gauge invariant.  In the doped case the same result is less obvious
since the spin particle can propagate by an exchange with a holon (or
doublon) and such a hopping process does not ``see'' the flux.  {\it
However}, the holon (or doublon) must move on a closed path and in the
end the $(-1)^{L_{nm}}$ factors involved in closing the path
compensate.  Similarly for charge propagators, e.g., within the
$t-J$-model $\langle T_{\tau}c^{\dagger}_{n\downarrow}(\tau)
c^{}_{m\downarrow}(0)\rangle = (-1)^{L_{nm}} \langle
T_{\tau}b^{}_{n}(\tau) b^{\dagger}_{m}(0)\rangle$ and similar
statements can be made.  Thus in general, e.g.,
$$
\langle [S^{+}_{n}(t),S^{-}_{m}(0)]\rangle = 
\exp[- i 
\int_{n}^{m} \vec a \cdot d \vec r]
\langle[f^{\dagger}_{n}(t),f^{}_{m}(0)]\rangle.
\eqno(13)
$$
When making potential gauge transformations it is worth noting that
the sense of a unite flux tube is immaterial.  In particular, for a
bipartite lattice, alternating the sense of the flux according to the
sub-lattice is an easy way to avoid making gauge transformations which
create (unphysical) currents at the boundaries.

Important also is the observation that, with the JW gauges, all matrix
elements of $\cal H$ are real.  For such a Hamiltonian matrix it is a
trivial fact that either the ground state vector is (i) real (or can
be made real with a simple change of phase) and therefore carries not
currents or (ii) is degenerate.  In case (ii) the real and imaginary
parts of the ground state vector correspond to degenerate states which
also carry no current.  Case (ii) does {\it not\/} exclude the
possibility of broken symmetry ground states which carry currents.  If
the degeneracy of the ground state is associated with the spin part of
the wave function then there must be a finite value of $S(S+1) = {\hat
S}^{2}$, i.e., such degenerate ground states must have a net
ferromagnetic moment, albeit as small as $S=1/2$.

Up to this point the flux tubes have been attached to the particles. 
It is equally possible to attach the flux tubes, and the false charge
which sees them, to the ``empty'', i.e., down spin sites.  This scheme
will be assumed in what follows.

It is clearly possible to associate an operator with a flux tube at a
given site.  In the operator $u^{}_{n}(\{\hat Q_{i}\}) = \exp[- i
\int_{0}^{n-1} \vec a \cdot d \vec r]$ the integral from the origin to
site $n$ passes by any path which does not include particles and
specifically along some path which does not pass through any sites. 
The argument $\{\hat Q_{i}\}$ indicates that $u^{}_{n}$ is a function
of the position of all of the particles, or more precisely reflects
the position of all of the down spin sites since it is these which
determine $\vec a$.  Clearly this is a unitary operator such that
$u^{\dagger}_{n} u^{}_{n} = u^{}_{n} u^{\dagger}_{n} = 1$.  For a JW
gauge $u^{\dagger}_{n} = u^{}_{n}$.  Trivially these $u^{\dagger}_{n}$
operators commute with each other but anti-commute with the particle
operators.  This formal development permits the, e.g., $t_{ij}
b^{\dagger}_{i} b^{}_{j}$ term to be written as:
$$
- t  b^{\dagger}_{i} u^{}_{i}  u^{\dagger}_{j} b^{}_{j}.
\eqno(14)
$$
and e.g., $c^{\dagger}_{i\downarrow} = u^{\dagger}_{i} b^{}_{i}$.  It
should be noted that these unitary operators always appear in such a
fashion that $u^{\dagger}_{i}$ is associated with the creation of an
down spin site {\it along with a flux tube\/} while $u^{}_{i}$ only
occurs when such a tube is destroyed.  Using these operators the 
$t-J$ model,
$$
{\cal H} = 
- t \sum_{<ij>}  ( f^{\dagger}_{i} b^{}_{i} b^{\dagger}_{j} f^{}_{j} 
+  b^{\dagger}_{i} u^{}_{i}  u^{\dagger}_{j} b^{}_{j} + H.c.)
+  \sum_{<ij>}{J\over 2}
(f^{\dagger}_{i}u^{}_{i}  u^{\dagger}_{j} f^{}_{j} + H.c.)
$$
$$
+ {J\over 4}\sum_{<ij>}
(2f^{\dagger}_{i} f^{}_{i} + b^{\dagger}_{i} b^{}_{i})
(2 f^{\dagger}_{j} f^{}_{j}+ b^{\dagger}_{j} b^{}_{j})
- J (2 \hat N_{\uparrow}+ \hat N_{0}).
\eqno(15)
$$

\topinsert
\centerline{ \epsfxsize=5.truein \epsfbox{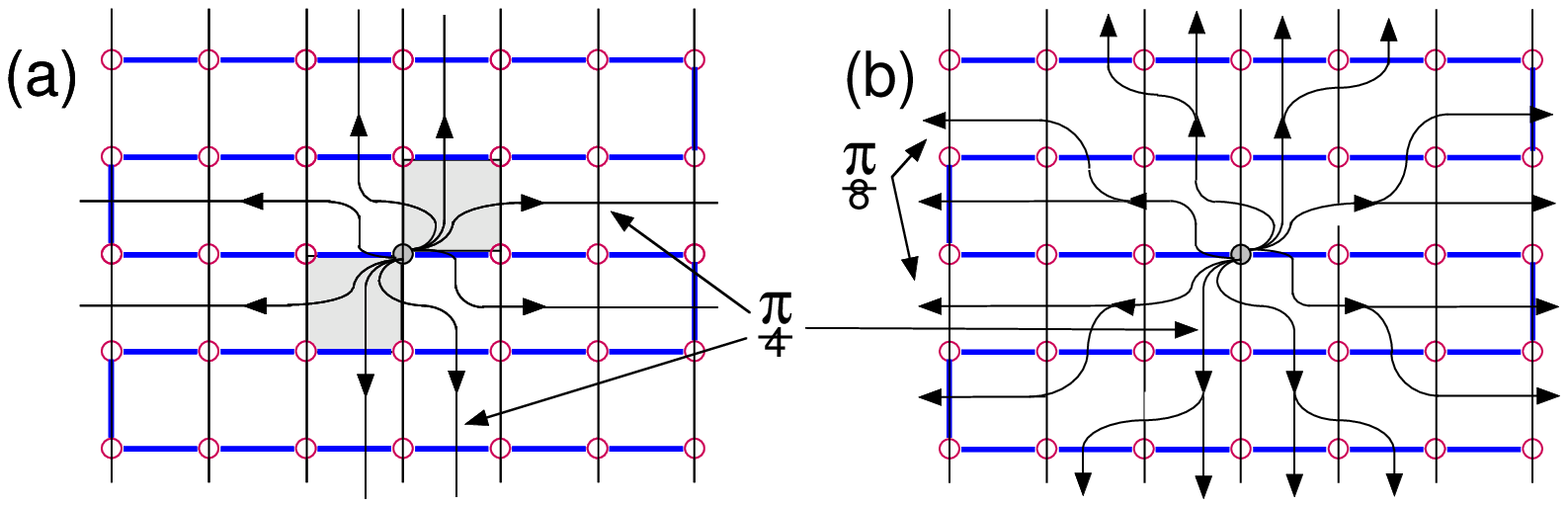} }
\noindent{Fig.  2 (a) The $\pi/4$ string gauge transformation.  There
is an effective half flux quantum in the shaded squares.  (b) The
equivalent with $\pi/8$ strings.  } 
\endinsert

The present JW-gauge does not reflect the symmetries of the lattice
and in addition the interaction between fermions which is induced by
the $\pi$-flux strings is of infinite range.  The usual gauge with
$\vec a = (1/r) \vec \phi$ is an evident alternative but is not
particularly adapted to lattice problems.  It is simpler to adapt a
different string gauge.  Each $\pi$ string can be thought of as made
of four $\pm \pi/4$ strings, where the sign is arbitrary.  Separating
one such string and moving it so it cuts different bonds, see
figure~2a, amounts to some gauge transformation.  (The flux pattern is
such that there is an effective half flux tube in each of the two
shaded plaquettes.)  If instead $\pi/8$ strings are used it is
possible to have the gauge convention shown in figure~2b.  Continuing
this bifurcation process will lead to interactions which $\sim (1/r)$. 
The arrow indicates the sign of the phase change.  As in
electrostatics, a positive magnetic flux tube with arrows heading away
from the tube will have positive sign changes when a particle crosses
a line heading in a clockwise direction.

\topinsert
\centerline{ \epsfxsize=5.truein \epsfbox{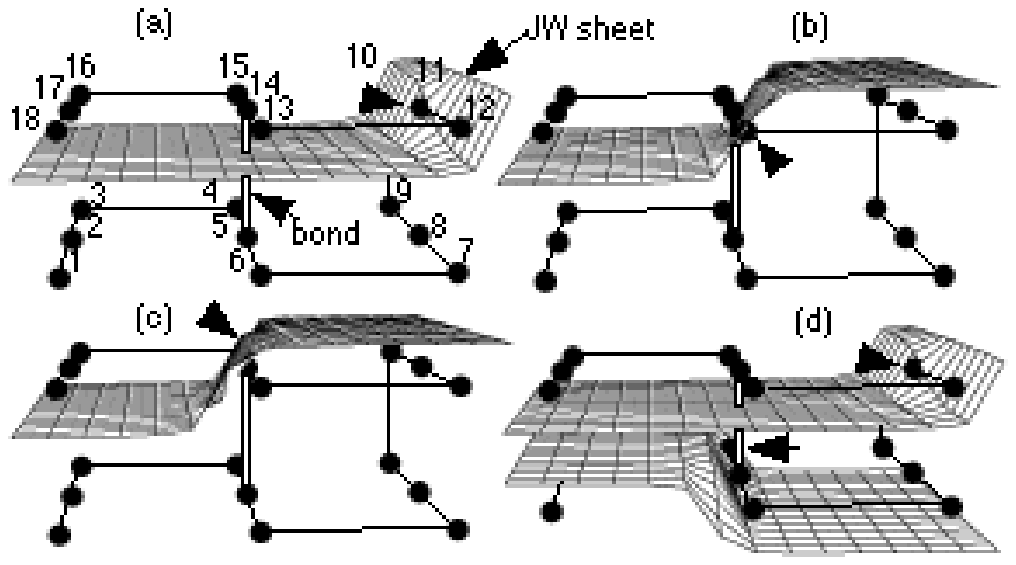} }
\noindent{Fig.  3 (a) A 18 site bi-layer is viewed as being to a
single folded plane and the zig-zag path $1, 2, \ldots, 17,18$ is
designated on that plane.  (Site 10 is hidden behind the sheet but can
be seen in (b) and (c).)  The JW-sheet is atatched to site 11 which
lies on the path between the illustrated bond between sites 5 and 14. 
That the sheet cuts this bond implies that the sign of the bond
changes.  (b) A similar JW-sheet attached to site 13 also lies on the
path between these sites and the sheet cuts the bond idicating again
that such a particle causes a change of sign.  However in (c) the
sheet is attached to site 15 is not on the path between 5 and 14 and
indeed the sheet does not cut the bond indicating that there is no
change in sign.  In general there are many sheets as in (d).  The
sheet attcahed to site 4 in the lower plane does not cut the bond and
there is therefore only a single change of sign associated with the
particle at site 11.  Explicity the sheet which passes throught site
11, coordinates $x=-1$, $y=0$, and $z=1$, was generated by $z = 1 -
0.4 \tan^{-1}[3(y+4\tan^{-1}\{3(x+1)\}]$.} \endinsert

It is of interest to generalise the present approach to bi-layers and
three dimensions in general.  While the fermion-plus-flux-tube type of
aynons do not generalise to three dimensions there is no real problem
with applying the JW transformation.  A bi-layer can be considered as
a single sheet which is folded over and the single sheet is then
reduced to one dimensions using again, e.g., the zig-zag path of
figure~1a.  Fully three dimensional system can similarly be mapped to
a corrugated sheet and then the sheet reduced to one dimension.  On
two dimensions the JW-transformation is effected by attaching $\pi$
flux strings to the particles.  The generalisation to three dimensions
is to attach sheets to each particle.  Consider, by way of
illustration the 18-site bi-layer shown in figure~3.  By design, if a
sheet is cut at the level of a given plane then the resulting line has
the same topology as the flux strings of figure~1 and the sign of a
intra-plane bond (not shown) is given, as before, by $(-1)^{L}$ where
$L$ is the number of such lines, or now sheets, which cross a bond. 
Each site on the zig-zag path, indicated by the black lines, has be
labled and advances in the sense $1, 2, \ldots, 17,18$ as shown in
figure~3a.  Focus attention on the white bond which connects sites 5
and 14.  Within the JW transfromation the sign associated with this
bond is $(-1)^{S}$ where $S$ is the number of particles on the sites
$6,7, \ldots, 12,13$ which lie between the ends of the bond on the
prescribed path.  The number $S$ is counted using the JW-sheets.  In
figure~3a the sheet is attached to a particle at site 11, identified
by an arrow.  This sheet cuts the bond indicating, correctly, that
this site must be included in $S$.  Similarly, figure~3b, the sheet
attached to site 13 cuts the bond and is to be reflected in $S$. 
However site 15 which lies on the same row as this latter site should
not have an effect on $S$ and indeed, figure~3c, the sheet attached to
this site does not cut the bond.  Of course, in general there are many
particles and many sheets which potentially are reflected in the factor
$(-1)^{S}$.  In figure~3d are shown two sheets, the one attached to
site 4 is not in the sequence $6,7, \ldots, 12,13$ and the sheet does
not cut the bond between 5 and 14, while, as described above that
attached to the site at 11 does, and should, cut the bond. Notice the 
sheets never cross the zig-zag path.

When two particles are interchanged it is necessarily the case that
one particle hops through the JW-sheet of the other particle an odd
number of times thereby converting the fermions to hard core bosons. 
Once defined with a specific zig-zag path,  deforming the sheets
amounts to a gauge transformation, and a $\pi$ sheet can be separated
into $n$, $\pi/n$ sheets which can be deformed almost at will.  In
this way the interactions between particles, reflected by the sheets,
can be made short ranged.  (Also, clearly, substituting a
$\theta$-sheet for the $\pi$-sheet permits an extrapolation between
fermions and hard core bosons in three dimensions.)

Whatever the dimension, the {\it bare\/} vacuum $|\rangle_{- N/2}$
reflects the absence of particles and comprises the spin state with
$S_{z} = - N/2$.  It is important that the bare ferromagnetic vacuum
state is highly degenerate.  A new vacuum $|\rangle_{- N/2+1} =
(S^{+}_{0}/M_{- N/2}^{- N/2})|\rangle_{- N/2+1}$ is obtained by acting
with the total spin raising operator $S^{+}_{0}$ where $M_{n}^{m} =
\sqrt{S(S+1) - nm}$.  The {\it physical\/} spin vacuum $|S\rangle$
comprises the Fermi spin sea.  To be specific define this as,
$$
|S\rangle 
=
\prod_{\epsilon_{\vec  k} \le 0} f^{\dagger}_{\vec k} |\rangle_{- 
N/2},
\eqno(16)
$$
where $\epsilon_{\vec k} \propto - \gamma_{\vec k}$; $\gamma_{\vec k}
= \cos k_{x}+ \cos k_{y}$.  Given that there are $N/2$, $f$-particles
(and that this is an integer) this is a spin {\it singlet}.  Although
this description of $|S\rangle$ is appealing, it is sometimes more
appropriate to imagine $|S\rangle$ as derived from the $S_{z}=0$ bare
vacuum $|\rangle_{0}^{\phantom{X}}$.  In the wave function $f^{\dagger}_{\vec k}$ is
replaced by $p_{\uparrow}=\sqrt{2} ( {1\over 2} + S_{n\, z} )$ and the
empty (down) sites are projected out using $p_{\downarrow}=\sqrt{2} (
{1\over 2} - S_{n\, z} )$.  With this $|S\rangle = F(\{S_{n\, z}\})
|\rangle_{0}^{}$, where $\{S_{n\, z}\}$ is the collection of local
$z$-component spin operators.

The states $f^{\dagger}_{\vec k} |S\rangle $ and $f^{}_{-\vec k}
|S\rangle $ are momentum $\vec k$ excitations with $S_{z} = \pm 1$
respectively and are related by a particle-hole symmetry at
half-filling.  In addition to these $f$-particle excitations, are
similar states generated by the flux tube operators $u^{}_{i}$.  A
arbitrary state with a given vacuum $S_{z}=n$ is of the form
$F_{1}(\{S_{n\, z}\})|\rangle_{n}^{}$ and an typical matrix element of
$u^{}_{i}$ is
$$
{}_{m}^{}\hskip - 1pt \langle |F^{*}_{2}u^{}_{i} F_{1}|\rangle_{n}^{} 
= 
{}_{m}^{}\hskip - 1pt \langle |F^{*}_{2}u^{}_{i}
( f^{\dagger}_{i} f^{}_{i} + f^{}_{i}f^{\dagger}_{i})
F_{1}|\rangle_{n}^{} 
$$
$$
\Big[
{}_{n}^{}\hskip - 1pt \langle |F^{*}_{2}f^{}_{i} F_{1}|\rangle_{n+1}^{} \delta_{m,n} 
+
{}_{n}^{}\hskip - 1pt \langle |F^{*}_{2}f^{\dagger}_{i} 
F_{1}|\rangle_{n-1}^{} \delta_{m,n}
$$
$$
\hskip 40pt+
{}_{n+1}^{}\hskip - 1pt \langle |F^{*}_{2}f^{\dagger}_{i} 
F_{1}|\rangle_{n}^{} \delta_{m,n+1}
+
{}_{n-1}^{}\hskip - 1pt \langle |F^{*}_{2} 
f^{}_{i} F_{1}|\rangle_{n}^{} \delta_{m,n+1}
\Big] + \ldots,
\eqno(17)
$$
where the ellipsis reflects terms which are vacuum on-diagonal.  Use
is made of the fact that both $u^{}_{i}$ and the projectors
$p_{\uparrow} = f^{\dagger}_{i} f^{}_{i}$ and $p_{\downarrow} =
f^{}_{i}f^{\dagger}_{i}$ commute with the $F_{n}(\{S_{n\, z}\})$ and
each other, and that e.g., $u^{}_{i} f^{\dagger}_{i} f^{}_{i} =
N^{-3/2}\sum_{\vec k \vec k^{\prime} \vec k^{\prime\prime} } e^{i\vec
k \cdot \vec r_{n}} e^{i\vec k^{\prime} \cdot \vec r_{n}} e^{i\vec
k^{\prime\prime} \cdot \vec r_{n}} u^{}_{\vec k} f^{\dagger}_{\vec
k^{\prime}}f^{}_{\vec k^{\prime\prime} } $.  The vacuum off-diagonal
part corresponds to contracting out a factor of $S^{\pm}/N$.  This
occurs in two ways for each term to give (17).  The two vacuum
diagonal terms have a different number of particles in the final
state, corresponding to (spin) hole and and particle excitations.  The
result can be written symbolically as
$$
u^{}_{i} ={1\over 2}
[f^{}_{i}S^{+} + f^{\dagger}_{i}S^{-} +
S^{+}f^{}_{i} + S^{-}f^{\dagger}_{i}] + \ldots,
\eqno(18)
$$
with the understanding that for, e.g., in $S^{+}f^{}_{i}$ the $S^{+}$ 
acts only on the $|\rangle_{0}$ vacuum to the left.
The ellipsis represents terms which are necessarily diagonal in the
bare vacuum state.  Since $u^{}_{n}$ is unitary $u^{}_{n} |S\rangle$
is a normalised state vector and it is to be noted that the four
vacuum off-diagonal terms displayed above exhausted exactly one half
of this normalisation.  The on-diagonal ellipsis, evidently, has equal
weight.

That, when it acts on the physical vacuum, the flux tube operator
$u^{}_{n}$ can create excitations (with a cancelling change in the
total $S_{z}$) is also evident from a comparison between $S_{z\, n} +
1/2 = f^{\dagger}_{n} f^{}_{n} $ and $S^{+}_{n} = f^{\dagger}_{n}
u^{}_{n}$.  The effect of these two operators acting upon the singlet
physical vacuum must be essentially the same, i.e., the effect of
$u^{}_{n}$ is the same as $f^{}_{n}$ {\it but\/} without a change in
$S_{z}$.  Multiplying (18) by $f^{+}_{n}$ gives, $S^{+}_{n} =
f^{\dagger}_{n} f^{}_{n}(S^{+}/N) + (S^{+}/N)
f^{}_{n}f^{\dagger}_{n}$.  (There is no commutator with $S^{+}$ in the
last term because of the understanding that this acts on the vacuum to
the left.)

Upon examining the ensemble of physical operators it should be
observed that these can be consistently interpreted in terms of
$f$-particles which are $S=1/2$ {\it spinons\/} with $S_{z} = +1/2$
while the flux-tube-particles are the equivalent with $S_{z} = -1/2$.

For the $t-J$ model, in the absence of doping, the spin sector kinetic
energy is generated by the vacuum diagonal part of the transverse
exchange,  i.e.,
$$
+ {J \over 2} f^{\dagger}_{i} u^{}_{i}  u^{\dagger}_{j} f^{}_{j}.
\eqno(19)
$$
For a near uniform state this must have an expansion $+ {J \over 2}
f^{\dagger}_{i} u^{}_{i} u^{\dagger}_{j} f^{}_{j} = a^{1}
f^{\dagger}_{i} f^{}_{j} + f^{\dagger}_{i} f^{}_{j} \sum a_{m}^{2}
\delta n_{m}$ where the $\delta n_{m}$ are deviations from the uniform
state.  Since, by definition $a^{1} = \langle u^{}_{i} u^{\dagger}_{j}
\rangle$ does not depend upon the uniform state involved, it can be
evaluated, with a suitable gauge, for a state vector which is an equal
weight of the two N\'eel states.  Both states have a uniform field
with one half a flux quantum per plaquette and in a (non-JW) gauge
which reflects the symmetry of the lattice this implies that the
change of phase introduced by the $u^{}_{i} u^{\dagger}_{j} \to e^{\pm
i\pi/4}$.  There are no direct matrix elements between these two state
and so $a^{1}$ is given by a simple average of these two phases, i.e.,
the leading term site diagonal part of (19) is,
$$
+ {J \over 2} \cos {\pi \over 4} f^{\dagger}_{i}  f^{}_{j} \equiv 
 { J^{\prime}\over 2} f^{\dagger}_{i}  f^{}_{j},
\eqno(20)
$$
with $J^{\prime} = J/\sqrt{2}$. Thus, e.g. for a two dimensions the 
bare energy of an excitation of momentum $\vec k$ is $\epsilon_{\vec 
k} = 2J^{\prime}\gamma_{\vec k}$.

The next term in the expansion of (19), written as $+{J \over 2}
S^{+}_{i}S^{-}_{j}$, is obtained using the vacuum off-diagonal part of
$S^{+}_{n}$ deduced above.  There are four equivalent contributions to
give the result,
$$
+ J  f^{\dagger}_{i} f^{}_{i} f^{}_{j} f^{\dagger}_{j} 
   + H.c. \Rightarrow
+ J  f^{\dagger}_{i} f^{\dagger}_{j} \langle f^{}_{j} f^{}_{i} 
\rangle + H.c. \equiv
\pm \Delta_{0} f^{\dagger}_{i} f^{\dagger}_{j} + H.c.,
\eqno(21)
$$
admits a pair amplitude.  With suitable signs, the Fourier transform
corresponds to $d$-wave pairing, i.e., a term,
$$
{1\over N} \sum_{\vec k}
\Delta_{\vec k} f^{\dagger}_{\vec k} f^{\dagger}_{-\vec k} + H.c.,
\eqno(22)
$$
where $\Delta_{\vec k} = \Delta_{0} (\cos k_{x} - \cos k_{y})$. 
Clearly, the original exchange term does not change the value of
$S_{z}$ and the pairing in (22) is to be interpreted in terms of
$\vec k$, $S_{z} = 1/2$ and $- \vec k$, $S_{z} = - 1/2$ spinon pairs.

The final version (15) of the JW formalism would appear very similar
to the usual auxiliary (slave boson) particle scheme.  The down spin
fermion has been replaced by a flux tube operator $u^{\dagger}_{i}$. 
However this parallel is {\it very\/} misleading.  In terms of up and
down fermions the transverse exchange is: $ + {J \over 2}\sum_{<ij>}
(f^{\dagger}_{\uparrow i} f^{}_{\downarrow i} f^{\dagger}_{\downarrow
j} f^{}_{\uparrow j} + H.c.)$.  Invariably in the ``slave boson''
approach this interaction is subjected to a mean field factorisation:
$$
 - \sum_{<ij>}{J \over 2}
 \langle f^{\dagger}_{\downarrow j} f^{}_{\downarrow i} \rangle
f^{\dagger}_{\uparrow i} f^{}_{\uparrow j} + H.c.,
\eqno(23)
$$
which changes the sign and reduces the magnitude of the matrix
elements for the hopping of the spin particles.  This latter
philosophy is very attractive since it separates the up and down
degrees of freedom and leads to an identification of the
$f^{\dagger}_{\sigma i}$ as spinons, i.e., spin $1/2$ particles.  The
cost is that the constraint is all but sacrificed.  The point of the
present formalism is precisely to maintain this constraint exactly and
yet to show that the excitations are indeed spinons.

The present formalism has its own naturel approximations.  If $t \gg
J$, the motion of the holes is certainly rapid compared to the
dynamics of the lowest lying spin levels.  If the concentration of
holes is sufficient there will occur a motional averaging of the spin
wave function.  The co-movement of the spins, as the holes hop, will
cause the spinon-spinon interactions to be averaged and suppress the
associated correlations (and ordering).  Given their rapid motion, the
holons can be added by perturbation theory.  On the charge time scale
the spins can be considered as frozen.  Consider a given up spin at
site $n$, it sees itself in a virtual crystal in which there is a
certain density of holes at near neighbour sites.  This up spin then
admixes with the holes on the neighbour sites and this implies a
probability $\sim x t/|\mu|\to x$ of observing a hole at $n$. 
Whatever the details, it is implied that the state which was
$f^{\dagger}_{n}|\rangle$ now becomes,
$$
(\alpha f^{\dagger}_{n}+ \beta
b^{\dagger}_{n} u^{\dagger}_{n} )|\rangle
\equiv \tilde f^{\dagger}_{n} |\rangle,
\eqno(24)
$$
where $|\alpha|^{2}+|\beta|^{2} =1$ with the concentration of holes $x
=|\beta|^{2}$ and where the flux tube $u^{\dagger}_{n} $ is needed in
order that $\tilde f^{\dagger}_{n}$ have fermion commutations rules. 
A Fermi sea is now constructed from the Fourier transform of the
$\tilde f^{\dagger}_{n} $.  The combination found in $\tilde
f^{\dagger}_{n}$ implies a pairing between the $b^{\dagger}_{n}$ {\it
holons\/} and $f^{}_{n}$.  This can be associated with Kondo physics
since with a small $\beta$ the excitations are relatively heavy. 
Likewise, the state for a down spin site,
$$
u^{\dagger}_{n}|\rangle \to 
(\alpha u^{\dagger}_{n} +\beta b^{\dagger}_{n} u^{\dagger}_{n} )|\rangle.
\eqno(25)
$$
With $S_{z}=0$ the presence of a $\tilde f^{\dagger}_{n}$ particle
implies a probability of $1-x$ of finding an up spin at site $n$, and
there are $N/2$ such particles independent of $x$.  In total, there
are $(N/2) (1+ x)$ degrees of freedom for such a system and the
present ``Kondo'' approximation implies that there are only $N/2$ low
energy energy excitations and that these are predominantly
``spinon-like'' with a small admixture of holon character.  For low
enough energies, the remaining holes are reflected, (25), as a
renormalisation of the vacuum.  The site independence of $\beta$
implies a holon {\it condensate\/} with $\vec q =0$.  The admixture of
holon character into the $f$-particles implies that the term, e.g., $-
t b^{\dagger}_{i} u^{}_{i} u^{\dagger}_{j} b^{}_{j}$ contributes to
the hopping matrix elements for these renormalisation particles.  As a
result $J^{\prime} \to J_{e} = (1-x) J^{\prime} - 2 xt$.  With this
approximation, the effective Hamiltonian is,
$$
{\cal H}_{s} = J_{e} \sum_{\vec k} \gamma_{\vec k}
\left(\tilde f^{\dagger}_{\vec k}\tilde f^{}_{\vec k} + H.c. \right) 
+ {(1-x )^{2}}
\sum_{\vec k}\left[ \Delta_{\vec k} \tilde f^{\dagger}_{\vec k}
\tilde f^{\dagger}_{-\vec k} + H.c.\right].  
\eqno(26)
$$
All interactions between spinons and holons which appeared in (15)
have been dropped since these will be average towards constants by the
rapid hole motion.  The physical conduction electron propagators
$\langle T_{\tau}c^{\dagger}_{n\uparrow}(\tau)
c^{}_{m\uparrow}(0)\rangle = c \langle T_{\tau}f^{}_{n}(\tau)
f^{\dagger}_{m}(0)\rangle$, i.e., are $c$, the condensed faction of
holons, times the spinon propagators.

Of considerable importance is the fact that the kinetic energy is
proportional to $J_{e} = \left( {J^{\prime}} -2 x t \right) $. 
Nagaoka's theorem [3] implies that a small hole doping favours
ferromagnetism.  The ferromagnetic, i.e., negative sign of the $2x t$
reflects this fact.  (This is the itinerant hole version of ring
exchange.)  Clearly the anti-ferromagnetic $J^{\prime}$-term must have
the opposite sign.  As a consequence, {\it in absence of pairing},
this Hamiltonian would exhibit a quantum critical point (QCP) at
$x_{c} = (J^{\prime}/2t)$ at which the spinon band collapses.  Clearly
the specific heat and magnetic susceptibility diverge at such a point. 
Since superconductivity is favoured by such a large density of states,
$T_{c}$ will be a maximum near $x_{c}$ and superconductivity will
``hide'' this QCP. This might be described as a pseudo-QCP with the
real QCP being shifted to smaller concentrations at which the holons
condensate and superconductivity first appears.

\def\refIII{
[3] 
Nagaoka Y 1965 Solid State Comm, {\bf 3} 409;
1966 Phys. Rev. {\bf 147} 392, see also: Shastry B S, H. R. 
Krishnamurthy H R, and Anderson P W 1990 Phys. Rev. B{\bf 41} 2375-2379 
}

This work was supported by a Grant-in-Aid for Scientific Research on
Priority Areas from the Ministry of Education, Science, Culture and
Technology of Japan, CREST. SEB is on sabbatical leave from the
Physics Department, University of Miami, Florida, USA and wishes to
thank the members of IMR for their kind hospitality.  SM acknowledges
support of the Humboldt Foundation. 

\vskip 20pt

\noindent{\bf References}

\vskip 20pt

\parindent=0pt
\refI

\refII

\refIII

\bye